\documentclass{article}
\usepackage{ijcai21}

\usepackage{times}

\usepackage{soul}
\usepackage{url}
\usepackage[hidelinks]{hyperref}
\usepackage[utf8]{inputenc}
\usepackage[small]{caption}
\usepackage{graphicx}
\usepackage{amsmath}
\usepackage{amsthm}
\usepackage{booktabs}
\usepackage{amsfonts}       
\usepackage{nicefrac}       
\usepackage{microtype}      
\usepackage{algorithm}
\usepackage{algorithmic}
\usepackage[ruled,linesnumbered,algo2e]{algorithm2e}

\usepackage{booktabs} 
\usepackage{amsmath,url,graphicx,amssymb}
\usepackage{amsfonts}
\usepackage{amsthm}
\usepackage{ctable}
\usepackage{multirow}
\usepackage{pifont}
\usepackage{color}
\usepackage{subfigure}
\usepackage{enumitem}
\usepackage{sidecap}
\usepackage{xcolor}
\usepackage[normalem]{ulem}

\newcommand{\rom}[1]{\uppercase\expandafter{\romannumeral #1\relax}}

\newcommand{\mylistbegin}{
  \begin{list}{$\bullet$}
   {
     \setlength{\itemsep}{-2pt}
     \setlength{\leftmargin}{1em}
     \setlength{\labelwidth}{1em}
     \setlength{\labelsep}{0.5em} } }
\newcommand{\mylistend}{
   \end{list}  }

\newcommand{\eg}{\textit{e.g.}}

\newcommand{\ie}{\textit{i.e.}}

\newcommand{\wrt}{\textit{w.r.t.~}}

\newcommand{\header}[1]{{\vspace{+1mm}\flushleft \textbf{#1}}}

\newtheorem{theorem}{Theorem}
\newtheorem{definition}{Definition}

\newtheorem{corollary}{Corollary}[theorem]

\newtheorem{thm:def}{Definition}
\newtheorem{thm:lemma}{Lemma}

\newcommand{\dgg}{\textsc{DPGGan}}
\newcommand{\back}{\textsc{DPGVae}}

\makeatletter
\newcommand{\nosemic}{\renewcommand{\@endalgocfline}{\relax}}

\title{Secure Deep Graph Generation with Link Differential Privacy}

\author{
  Carl Yang$^1$\footnote{Corresponding Author}\and
  Haonan Wang$^2$\and 
  Ke Zhang$^3$\and 
  Liang Chen$^4$\and 
  Lichao Sun$^5$\\
  \affiliations
  $^1$Emory University, 
  $^2$University of Illinois at Urbana Champaign\\
  $^3$University of Hong Kong,
  $^4$Sun Yat-sen University,
  $^5$Lehigh University
  \emails
  j.carlyang@emory.edu,
  haonan3@illinois.edu,
  cszhangk@connect.hku.hk\\
  chenliang6@mail.sysu.edu.cn,
  lis221@lehigh.edu
}

\begin{document}

\maketitle
\begin{abstract}
Many data mining and analytical tasks rely on the abstraction of networks (graphs) to summarize relational structures among individuals (nodes). Since relational data are often sensitive, we aim to seek effective approaches to generate utility-preserved yet privacy-protected structured data.
In this paper, we leverage the differential privacy (DP) framework to formulate and enforce rigorous privacy constraints on deep graph generation models, with a focus on edge-DP to guarantee individual link privacy.
In particular, we enforce edge-DP by injecting proper noise to the gradients of a link reconstruction based graph generation model, while ensuring data utility by improving structure learning with structure-oriented graph discrimination.
Extensive experiments on two real-world network datasets show that our proposed \dgg~model is able to generate graphs with effectively preserved global structure and rigorously protected individual link privacy.
\end{abstract}

\section{Introduction}
\label{sec:intro}
Nowadays, open data of networks (graphs) play a pivotal role in data mining and data analytics \cite{hu2020open,xie2020gnns}. 
By releasing and sharing structured relational data with research facilities and enterprise partners, data companies harvest the enormous potential value from their data, which benefits decision-making on various aspects, including social, financial, environmental, through collectively improved ads, recommendation, retention, and so on \cite{sigurbjornsson2008flickr,kuhn2009compensation}. 
However, graph data usually encode sensitive information not only about individuals but also their interactions, which makes direct release and exploitation rather unsafe. 
More importantly, even with careful anonymization, individual privacy is still at stake under collective attack models facilitated by the underlying graph structure \cite{zhang2019enabling,cai2018collective,sun2018adversarial}. 
Can we find a way to securely generate graph data without drastic sanitization that essentially renders the released data useless?

In dealing with such tension between the need to release utilizable data and the concern of data owners' privacy, 
quite a few secure deep generative models have been proposed recently, focusing on grid-based data like images, texts and gene sequences \cite{papernot2018scalable,sun2020ldp,sun2020federated}. 
However, none of the existing models can be directly applied to the network (graph) setting. 
While a secure generative model on grid-based data apparently aims to preserve high-level semantics (\eg, class distributions) and protect detailed training data (\eg, exact images or sentences), it remains obtuse what to be preserved and what to be protected for graph data, due to its modeling of complex interactive objects.

\begin{figure}[t!]
\centering
\vspace{-3mm}
\subfigure[Anonymized original net.]{
\includegraphics[width=0.2\textwidth]{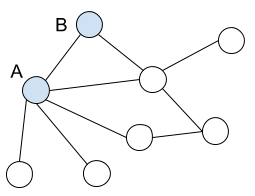}}
\subfigure[\dgg~generated net.]{
\includegraphics[width=0.2\textwidth]{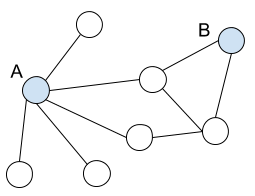}}
\vspace{-3mm}
\caption{\textbf{A toy pair of anonymized and generated networks.}}
\vspace{-4mm}
\label{fig:toy}
\end{figure}

\vspace{-2mm}
\header{Motivating scenario.} 
In Figure \ref{fig:toy}, a bank aims to encourage public studies on its customers' community structures. It does so by firstly anonymizing all customers and then sharing the network (\ie, (a) in Figure \ref{fig:toy}) to the public.
However, an attacker interested in knowing the financial interactions (\eg, money transfer) between particular customers in the bank may happen to have access to another network of a similar set of customers (\eg, a malicious employee of another financial company). 
The similarity of simple graph properties like node degree distribution and triangle count between the two networks can then be used to identify specific customers with high accuracy in the released network (\eg, customer $A$ as the only node with degree 5 and within 1 triangle, and customer $B$ as the only node with degree 2 and within 1 triangle). Thus, the attacker confidently knows the A and B's identities and the fact that they have financial interactions in the bank, which seriously harms customers' privacy and poses potential crises. 

As the first contribution in this work, we define and formulate the goals of secure graph generation as \textit{preserving global graph structure} while \textit{protecting individual link privacy}. Continue with the toy example, the solution we propose is to train a deep graph generation model on the original network and release the generated networks (\eg, (b) in Figure \ref{fig:toy}). 
Towards the utility of generated networks, we require them to be similar to the original networks from a global perspective, which can be measured by various commonly graph properties (\eg, network (b) has very similar degree distribution and the same triangle count as (a)). In this way, we expect many downstream data mining and analytical tasks on them to produce similar results as on the original networks.
As for privacy protection, we require that information in the generated networks cannot confidently reveal the existence or absence of any individual links in the original networks (\eg, the attacker may still identify customers A and B in network (b), but their link structure may have changed).

Subsequently, there are two unique challenges in learning such structure-preserved and privacy-protected graph generation models, which have never been explored so far.

{\flushleft\bf Challenge 1: Rigorous protection of individual link privacy.}
The rich relational structures in graph data often allow attackers to recover private information through various ways of collective inference \cite{zhang2014privacy,narayanan2009anonymizing}. 
Moreover, graph structure can always be converted to numerical features such as spectral embedding, 
after which most attacks on grid-based data like model inversion \cite{fredrikson2015model} and membership inference \cite{shokri2017membership} can be directly applied for link identification. 
Existing frameworks~\cite{nobari2014opacity,xue2012delineating} for graph link protection only demonstrate certain types of privacy regarding specific empirical measures without principled theoretical guarantee.
How can we design an effective mechanism with rigorous privacy protection on links in graphs against various attacks?

{\flushleft\bf Challenge 2: Effective preservation of global graph structure.}
To capture the global graph structure, the model has to constantly compare the structures of the input graphs and currently generated graphs during training.
However, unlike images and other grid-based data, graphs have flexible structures and arbitrary node orders.
How can we allow a graph generation model to effectively learn from the structural difference between two graphs, without conducting very time-costly operations like isomorphism tests all the time?

\vspace{-2mm}
\header{Present work.}
In this work, for the first time, we draw attention to the secure generation of graph data with deep neural networks. Technically, towards the aforementioned two challenges, we develop Differentially Private Graph Generative Adversarial Networks (\dgg), which imposes DP training over a link reconstruction based graph generation model for rigorous individual link privacy protection, and further ensures structure-oriented graph comparison for effective global graph structure preservation.
In particular, we first formulate and enforce edge-DP via gradient distortion by properly injecting designed noises during model training. Then we leverage graph convolutional networks \cite{kipf2016semi} through a generative adversarial network architecture \cite{gu2018dialogwae} to enable structure-oriented graph discrimination.

To evaluate the effectiveness of \dgg, we conduct extensive experiments on two real-world network datasets. 
On one hand, we evaluate the utility of generated graphs by computing a suite of commonly used graph properties to compare the global structures of generated graphs with the original ones. On the other hand, we validate the privacy of individual links by evaluating links predicted from the generated graphs on the original graphs. Consistent experimental results show that \dgg~is able to effectively generate graphs that are similar to the original ones regarding global graph structure, while at the same time useless towards individual link prediction. 

%
%

\section{Related Work}
\label{sec:related}

\header{Differential Privacy (DP).} Differential privacy is a statistical approach in addressing the paradox of learning nothing about an individual while learning useful information about a population~\cite{dwork2014algorithmic}. 
Recent advances in deep learning have led to the rapid development of DP-oriented learning schemes.
Among them, the Gaussian Mechanism \cite{dwork2014algorithmic}, defined as follows, provides a neat and compatible framework for DP analysis over machine learning models.
\begin{definition}[Gaussian Mechanism \cite{dwork2014algorithmic}]
\label{thm:gaussian}
For a deterministic function $f$ with its $\ell_2$-norm sensitivity as $\Delta _2 f=\max\limits_{\Vert \mathbf{G}-\mathbf{G}'\Vert_1=1}\Vert f(\mathbf{G})-f(\mathbf{G}')\Vert_2$, we have:
    \begin{equation}
    \mathcal{M}_f (\mathbf{G}) \triangleq f(\mathbf{G})+\mathcal{N}(0, {\Delta_2 f}^2\sigma^2),
    \label{eq:gauss}
    \end{equation}
where $\mathcal{N}(0, {\Delta _2 f}^2\sigma^2)$ is a random variable obeying the Gaussian distribution with mean 0 and standard deviation ${\Delta _2 f}\sigma$. The randomized mechanism $\mathcal{M}_f(\mathbf{G})$ is $(\varepsilon,\delta)$-DP if $\sigma  \ge \Delta _2 f\sqrt {2\ln (1.25/\delta )} /\varepsilon $ and $\varepsilon < 1$.
\end{definition}
Following this framework, \cite{Abadi:2016:DLD:2976749.2978318} proposes a general training strategy called DPSGD, which looses the condition on the overall privacy loss than that in Definition \ref{thm:gaussian} by tracking detailed information of the SGD process to achieve an adaptive Gaussian Mechanism. 

DP learning has also been widely adapted to generative models \cite{frigerio2019differentially,papernot2018scalable,sun2020differentially}.
For example, \cite{frigerio2019differentially} enforces DP on the discriminators, and thus inductively on the generators, in a GAN scheme.
However, none of them can be directly applied to graph data due to the lack of consideration of structure generation.

For graph structured data, two types of privacy constraints can be applied, \ie, node-DP \cite{Kasiviswanathan13NodeDP} and edge-DP \cite{Blocki12EdgeDP},
which define two neighboring graphs to differ by at most one node or edge. 
In this work, we aim at secure graph generation, and particularly, we focus on edge privacy because it is essential for the protection of object interactions unique for graph data.  
Several existing works have studied the protection of edge-DP. For example,
\cite{Sala11ShareGraphDP} generates graphs based on the statistical representations extracted from the original graphs blurred by designed noise, whereas \cite{Wang13DPDegreeGraphGeneration} enforces the parameters of dK-graph models to be private. 
However, based on shallow graph generation models, they do not flexibly capture global graph structure that can support various unknown downstream analytical tasks.

\vspace{-2mm}
\header{Graph Generation (GGen).}
GGen has been studied for decades and is widely used to synthesize network data for developing various collective analysis and mining models. 
Earlier works mainly use probabilistic models to generate graphs with certain properties \cite{watts1998collective,barabasi1999emergence}, 
which are manually designed based on sheer observations and prior assumptions. 

Thanks to the surge of deep learning, many advanced GGen models have been developed recently, which leverage different powerful neural networks in a learn-to-generate manner \cite{kipf2016variational,bojchevski2018netgan,you2018graphrnn,simonovsky2018graphvae} 
For example, 
NetGAN \cite{bojchevski2018netgan} converts graphs into biased random walks, learns the generation of walks with GAN, and assembles the generated walks into graphs; GraphRNN \cite{you2018graphrnn} regards the generation of graphs as node-and-edge addition sequences, and models it with a heuristic breadth-first-search scheme and hierarchical RNN. These deep learning models can often generate graphs with much richer properties, and flexible structures learned from real-world networks.

To the best of our knowledge, no existing work on deep GGen has looked into the potential privacy threats laid during the learning and generation with powerful models. Such concerns are rather urgent in the graph setting, where sensitive information can often be more easily compromised in a collective manner \cite{zhang2014privacy}, and privacy leakage can easily further propagate \cite{sun2018adversarial}. 

\section{\dgg}
\label{sec:model}
In this work, we propose \dgg~for securely generating graphs, whose global structures are similar to the original sensitive ones, but the individual links (edges) between objects (nodes) are safely protected. 
To provide robust privacy guarantees towards various graph attacks, we propose to leverage the well-studied technique of differential privacy (DP) \cite{dwork2014algorithmic} by enforcing the edge-DP defined as follows.

\begin{definition} [Edge Differential Privacy \cite{Blocki12EdgeDP}]
\label{def:EdgeDP}
A randomized mechanism $\mathcal{M}$ satisfies $(\varepsilon,\delta)$-edge-DP 
if for any two neighboring graphs $\mathbf{G}_1, \mathbf{G}_2 \in \mathcal{G}$, which differ by at most one edge,
$\Pr[\mathcal{M}(\mathbf{G}_1) \in S] \leq exp(\varepsilon) \times \Pr[\mathcal{M}(\mathbf{G}_2) \in S] +\delta$, where  $S \subset range(\mathcal{M})$. 
\end{definition}

Our key insight is, \textit{a graph generation model $\mathcal{M}$ satisfying the above edge-DP should learn to generate similar graphs given the input of two neighboring graphs that differ by at most one edge; as a consequence, the information in the generated graph does not confidently reveal the existence or absence of any one particular edge in the original graph, thus rigorously protecting individual link privacy.} 

To ensure DP on individual links, we exploit the existing link reconstruction based graph generation model GVAE \cite{kipf2016variational}, and design a training algorithm to dynamically distort the gradients of its sensitive model parameters by injecting proper amounts of Gaussian noise based on the framework of DPSGD \cite{Abadi:2016:DLD:2976749.2978318}.
We provide theoretical analysis on applying DPSGD to achieve edge-DP with GVAE based on the nature of graph data and the link reconstruction loss.
Moreover, to improve the capturing of global graph structures, we replace the direct binary cross-entropy (BCE) loss on graph adjacency matrices in GVAE with a structure-oriented graph discriminator based on GCN and the framework of VAEGAN \cite{gu2018dialogwae}.
We further prove the improved model to maintain the same edge-DP.



\header{Backbone GVAE.}
Recent research on graph models has been primarily focused around GCN \cite{kipf2016semi}, which is shown to be promising in calculating universal graph representations \cite{maron2019invariant,xu2018powerful}. To guarantee individual link privacy without severely damaging global network structure, in this work, we harness the power and simplicity of GCN under the consideration of edge-DP by adapting the link reconstruction based graph variational autoencoder (GVAE) \cite{kipf2016variational} as our backbone graph generation model. 

Notably, we are given a graph $\mathbf{G}=\{\mathbf{V}, \mathbf{E}\}$, where $\mathbf{V}$ is the set of $N$ nodes (vertices), and $\mathbf{E}$ is the set of $M$ links (edges), which can be further modeled by a binary adjacency matrix $\mathbf{A}$. 
As a common practice \cite{hamilton2017inductive}, we set the node features $\mathbf{X}$ simply as the one-hot node identity matrix. The autoencoder architecture of GVAE consists of a GCN-based graph encoder to guide the learning of a feedforward neural network (FNN) based adjacency matrix decoder, which can be trained to directly reconstruct graphs with similar links as in the input graphs. A stochastic latent variable $\mathbf{Z}$ is further introduced as the latent representation of $\mathbf{A}$ as
\begin{align}
q(\mathbf{Z} | \mathbf{X}, \mathbf{A}) = \prod_{i=1}^N q(\mathbf{Z}_i|\mathbf{X}, \mathbf{A}) =\prod_i^N \mathcal{N}(\mathbf{z}_i|\mathbf{\mu}_i, \text{diag}(\mathbf{\sigma}_i^2)), 
\label{eq:encode}
\end{align}
where $\mathbf{\mu}=\mathbf{g}_\mu(\mathbf{X}, \mathbf{A})$ is the matrix of mean vectors $\mathbf{\mu}_i$, and $\mathbf{\sigma}=\mathbf{g}_\sigma(\mathbf{X}, \mathbf{A})$ is the matrix of standard deviation vectors $\mathbf{\sigma}_i$. 
$\mathbf{g}_{\bullet}(\mathbf{X}, \mathbf{A})= \tilde{\mathbf{A}}\text{ReLU}(\tilde{\mathbf{A}}\mathbf{X}\mathbf{W}_0)\mathbf{W}_1$ is a two-layer GCN model. 
$\mathbf{g}_\mathbf{\mu}$ and $\mathbf{g}_\mathbf{\sigma}$ share the first-layer parameters $\mathbf{W}_0$. $\tilde{\mathbf{A}}=\mathbf{D}^{-\frac{1}{2}}\mathbf{A}\mathbf{D}^{-\frac{1}{2}}$ is the symmetrically normalized adjacency matrix of $\mathbf{G}$, with degree matrix $\mathbf{D}_{ii}=\sum_{j=1}^N \mathbf{A}_{ij}$. $\mathbf{g}_\mu$ and $\mathbf{g}_\sigma$ form the encoder network. 

To generate a graph $\mathbf{G}'$, a reconstructed adjacency matrix $\mathbf{A}'$ is computed from $\mathbf{Z}$ by an FNN decoder
\begin{align}
p(\mathbf{A}|\mathbf{Z}) = \prod_{i=1}^N \prod_{j=1}^N p(\mathbf{A}_{ij}|\mathbf{z}_i, \mathbf{z}_j) = \prod_{i=1}^N \prod_{j=1}^N\sigma(\mathbf{f}(\mathbf{z}_i)^T\mathbf{f}(\mathbf{z}_j)),
\label{eq:decode}
\end{align}
where $\sigma(z) = 1/(1+e^{-z})$, $\mathbf{f}$ is a two-layer FNN appended to $\mathbf{Z}$ before the logistic sigmoid function. 
The whole model is trained through optimizing the following variational lower bound
\begin{align}
\mathcal{L}_{vae} &= \mathcal{L}_{\text{rec}} + \mathcal{L}_{\text{prior}} \label{eq:loss}\\
&= \mathbb{E}_{q(\mathbf{Z}|\mathbf{X},\mathbf{A})}[\log p(\mathbf{A}|\mathbf{Z})] - D_{\text{KL}}(q(\mathbf{Z}|\mathbf{X}, \mathbf{A})\Vert p(\mathbf{Z})),\nonumber
\end{align}
where $\mathcal{L}_{\text{rec}}$ is implemented as the sum of an element-wise binary cross entropy (BCE) loss between the adjacency matrices of the input and generated graphs, and $\mathcal{L}_{\text{prior}}$ is a prior loss based on the Kullback-Leibler divergence towards the Gaussian prior $p(\mathbf{Z}) = \prod_{i=1}^N p(\mathbf{z}_i)=\prod_i^N\mathcal{N}(\mathbf{z}_i|\mathbf{0}, \mathbf{I})$. 

\header{Enforcing DP.}
The probabilistic nature of $\mathbf{Z}$ allows the model to be generative, meaning that after training the model with an input graph $\mathbf{G}$, we can detach and disregard the encoder, and then freely generate an unlimited amount of graphs $\mathbf{G}'$ with similar links to $\mathbf{G}$, by solely drawing random samples of $\mathbf{Z}$ from the prior distribution $\mathcal{N}(\mathbf{0}, \mathbf{I})$ and computing $\mathbf{A}'$ with the learned decoder network \wrt~Eq.~\eqref{eq:decode}. 
However, as shown in \cite{kurakin2016adversarial}, powerful neural network models like VAE can easily overfit training data, so directly releasing a trained GVAE model poses potential privacy threats.

In this work, we care about the generation model's rigorously protecting the privacy of individual links in the training data, \ie, ensuring edge-DP.
Particularly, in Definition \ref{def:EdgeDP}, the inequality guarantees that the distinguishability of any one edge in the graph will be restricted to the privacy leak level proportional to $\varepsilon$, quantifying the absolute value of privacy information possibly to be leaked by a graph generation model.

According to Eq.~\eqref{eq:decode}, GVAE essentially takes a graph $\mathbf{G}$, in particular, the links $\mathbf{E}$ among the nodes $\mathbf{V}$ in $\mathbf{G}$, as input and generates a new graph $\mathbf{G}'$ by reconstructing the links $\mathbf{E}'$ among the same set of nodes $\mathbf{V}$. Therefore, if we regard GVAE as the mechanism $\mathcal{M}$, as long as its model parameters are properly randomized, the framework satisfies edge-DP. To be specific, any two input graphs $\mathbf{G}_1$ and $\mathbf{G}_2$ differing by at most one link in principle lead to similar generated graphs $\mathbf{G}'$, so information in $\mathbf{G}'$ does not confidently reveal the existence or absence of any particular link in $\mathbf{G}_1$ or $\mathbf{G}_2$. 

To exploit the well-structured graph generation framework of GVAE, we leverage the Gaussian mechanism (Definition \ref{thm:gaussian}) \cite{dwork2014algorithmic} and DPSGD \cite{Abadi:2016:DLD:2976749.2978318} to enforce edge-DP on it. 
In our setting, $\mathbf{G}$ is the original training graph. Then Eq.~\eqref{eq:gauss} tells us that a link reconstruction based graph generation model $\mathcal{M}$ can be randomized to ensure $(\varepsilon, \delta)$-edge-DP with properly parameterized Gaussian noise. 
Prominently, we follow DPSGD~\cite{Abadi:2016:DLD:2976749.2978318} to inject a designed Gaussian noise to the gradients of our decoder network clipped by a hyper-parameter C as follows
{\small
\begin{equation}
\tilde{g}_{\theta, \mathcal{L}}= \frac{1}{N} \left(\sum_{i=1}^N \left(\nabla_{v_i, \theta}\mathcal{L}/\text{max}(1, \frac{\Vert\nabla_{v_i, \theta}\mathcal{L}\Vert_2}{C})\right)+\mathcal{N}(0,\sigma^2C^2\bold{I})\right),
\label{eq:dpsgd}
\end{equation}}
where $\mathcal{L}$ is the loss function of a link reconstruction based graph generation model, $C$ is the clipping hyper-parameter for the model's original gradient to bound the influence of each link, and $\sigma$ is the noise scale hyper-parameter. 

Now we prove that the noised clipped gradient $\tilde{g}_{\theta, \mathcal{L}}$ applied as above guarantees the learned graph generation model to be edge-DP, with a different condition from that in Definition \ref{thm:gaussian} due to the nature of graph generation.

\begin{theorem}
\label{thm:dpsgd} In training a link reconstruction based graph generation model on a graph with $N$ nodes with batch size $B$, given the sampling probability $q=B/N$, and the number of steps $T$, there exist explicit constants $c_1$ and $c_2$ that for any $\varepsilon<c_{1} q^{2}T$, iteratively updating the model $T$ times with $\tilde{g}_{\theta, \mathcal{L}}$ attains it with $(\varepsilon, \delta)$-edge-DP for any $\delta>0$ if we choose
\begin{equation}
\nonumber
\sigma \geq c_{2} \frac{q \sqrt{T \log (1 / \delta)}}{\varepsilon},
\label{eq:noise}
\end{equation}
where $c_1\geq\frac{1}{c_0}\log{\frac{1}{q\sigma}}$, $c_2\leq1/\sqrt{c_0(1-c_0)}$ for any $c_0\in(0,1)$.
\vspace{-1mm}
\end{theorem}
The proofs of Theorem~\ref{thm:dpsgd} are detailed in Appendix A.

For the training of the \back~decoder, $\mathcal{L}$ in Eq.~\eqref{eq:dpsgd} is specified as $\mathcal{L}_{rec}$ in Eq.~\eqref{eq:loss}. Due to the link reconstruction nature of \back, we derive the following Corollary.

\begin{corollary}
	[\back~edge-DP]
	\label{thm:dpgvae}
	Under the same conditions in Theorem \ref{thm:dpsgd}, iteratively updating the decoder in \back~for $T$ times with $\tilde{g}_{\theta, \mathcal{L}_{rec}}$ attains it with $(\varepsilon, \delta)$-edge-DP.
	\end{corollary}

In the generation stage, we can disregard the encoder and only use the decoder to generate an unlimited amount of graphs from randomly sampled vectors from the prior distribution $\mathcal{N}(\mathbf{0},\mathbf{I})$. 
Due to the randomness of the normal Gaussian distribution, the sampling process can be regarded as $(0,0)$-DP. 
By the composability property of DP \cite{dwork2014algorithmic}, generating graphs from random noises with the \back~decoder satisfies $(\varepsilon, \delta)$-edge-DP,
whose release in principle does not disclose sensitive information regarding individual links in the original sensitive networks. 
Since we do not release the encoder network, we do not need to clip and perturb its gradients during training to induce minimum interruptions.

\begin{figure*}[t]
\centering
\vspace{-15pt}
\includegraphics[width=0.65\textwidth]{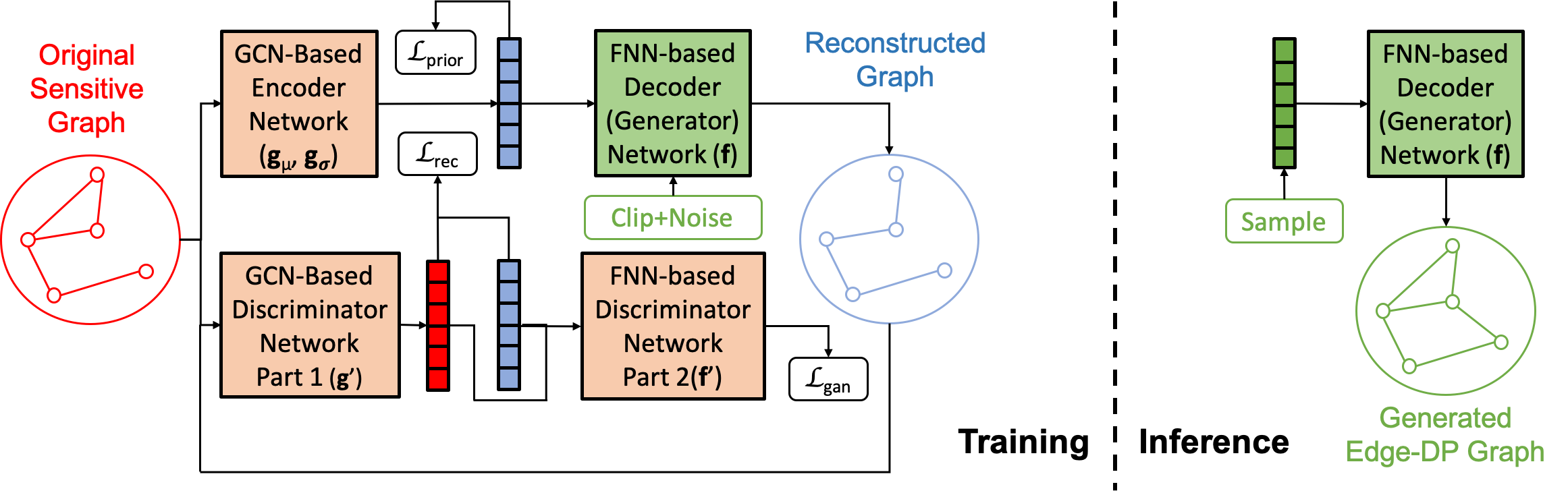}
\vspace{-5pt}
\caption{\textbf{Neural architecture of \dgg~(best viewed in color)}: Our novel graph generation model consists of a GCN-based encoder, an FNN-based decoder (generator), and a GCN+FNN-based discriminator. Sensitive data and modules are marked as red, while safe operations (\ie, gradient clipping, noise injection and sampling) are marked as green, leading to DP modules and data.}
\label{fig:arch}
\vspace{-15pt}
 \end{figure*}
 
\header{Improving structure learning.}
Besides individual link privacy, we also aim to preserve the global network structure to ensure the utility of released data. As we discuss before, original GVAE computes the reconstruction loss between input and generated graphs based on the element-wise BCE between their adjacency matrices. Such a computation is specified on each link, rather than the graph structure as a whole. To improve the global graph structure learning, we leverage GCN again, which has been shown universally powerful in capturing graph-level structures~\cite{maron2019invariant,xu2018powerful}. 
Therefore, we borrow the framework of VAEGAN and compute a structure-oriented generative adversarial network (GAN) loss as
\begin{align}
\mathcal{L}_{gan} = \log (\mathcal{D}(\mathbf{A})) + \log (1-\mathcal{D}(\mathbf{A}'))&\nonumber\\
\text{with} \; \mathcal{D}(\mathbf{A}) = \mathbf{f}'(\mathbf{g}'(\mathbf{X}, \mathbf{A})), &
\label{eq:loss2}
\end{align}
where $\mathbf{g}'$ and $\mathbf{f}'$ are GCN and FNN networks similar as defined before, besides at the end of $\mathbf{g}'$ the node-level representations are summed up as the graph-level representation, which resembles the recently proposed GIN model for graph-level representation learning \cite{xu2018powerful}. In this \dgg~framework, the decoder also serves as the generator, while $\mathcal{D}=\mathbf{f}'\cdot\mathbf{g}'$ is the discriminator.

Following \cite{gu2018dialogwae}, the encoder is trained \wrt~$\mathcal{L}_{rec}+\lambda_1\mathcal{L}_{prior}$, the generator  \wrt~$\mathcal{L}_{rec}-\lambda_2\mathcal{L}_{gan}$, and the discriminator \wrt~$\lambda_2\mathcal{L}_{gan}$, where $\lambda_1$ and $\lambda_2$ are hyper-parameters. 
In practice, to apply DPSGD to our graph generation model, we consider various options for noise injection towards the appropriate DP constraints, such as adding noises to the GCN encoder, latent graph representation, FNN decoder, GCN discriminator or any of their combinations. With theoretical justification and empirical analysis, we find injecting a designed Gaussian noise to the clipped gradients of our decoder network sufficiently leads to edge-DP of generated graphs while best preserving the desired global graph structures.
Therefore, Eq.~\eqref{eq:dpsgd} with $\mathcal{L}_{rec}$ substituted by $\mathcal{L}_{rec}-\lambda_2\mathcal{L}_{gan}$ is applied to distort the gradients of the decoder (generator) and guarantee edge-DP, which can be used to securely generate networks with the other parts disregarded after training.
Furthermore, to achieve the best performance without losing the DP guarantee, we gradually reduce $C$ in Eq.~\eqref{eq:dpsgd} for an adaptive perturbation.
The overall framework of {\dgg} is shown in Figure \ref{fig:arch}, and the training process is detailed in Appendix B.

The intuition behind the novel design of {\dgg} is, the GCN encodings $\mathbf{g}'(\mathbf{A})$ and $\mathbf{g}'(\mathbf{A}')$ capture the graph structures of $\mathbf{G}$ and $\mathbf{G}'$, so a reconstruction loss $\mathcal{L}_{rec}=\Vert\mathbf{g}'(\mathbf{A})-\mathbf{g}'(\mathbf{A}')\Vert^2_2$ captures the intrinsic structural difference between $\mathbf{G}$ and $\mathbf{G}'$ instead of the simple sum of the differences over their individual links. 
Note that the effectiveness of our structure-oriented discriminator is critical not only because it can directly enforce effective training of the graph generator through the minimax game in Eq.~\eqref{eq:loss2}, but also because it can learn to relax the penalty on certain individual links through flexible and diverse configurations of the whole graph as long as the global structures remain similar, which exactly fulfills our goals of secure network release. The benefits of such diversity enabled by the VAEGAN have also been discussed in image generation \cite{gu2018dialogwae}.

Compared with \back, \dgg~does not directly compute the link reconstruction loss based on BCE in Eq.~\eqref{eq:loss}, but rather computes it based on the graph discriminator $\mathcal{D}$. However, the link reconstruction based graph generator of \dgg~is exactly the same as \back. Since we also do not release $\mathcal{D}$ after training, we can simply retrieve Corollary~\ref{thm:dpggan} from Theorem \ref{thm:dpsgd} as follows.
\begin{corollary}
[\dgg~edge-DP]
\label{thm:dpggan}
Under the same conditions in Theorem \ref{thm:dpsgd}, iteratively updating the generator in \dgg~for $T$ times with $\tilde{g}_{\theta, (\mathcal{L}_{rec}-\lambda_2\mathcal{L}_{gan})}$ attains it with $(\varepsilon, \delta)$-edge-DP.
	\end{corollary} 

With Corollary \ref{thm:dpggan}, we attain \dgg~with the same $(\varepsilon, \delta)$-edge-DP protection of \back. For both \back~and \dgg, the decoder/generator networks only get exposed to the noised and clipped gradients, representing the partial sensitive information within the training graphs. Hence, it prevents the inference of training graphs from both learned model parameters and generated graphs. 

\section {Experimental Evaluations} 
\label{sec:exp}

\begin{table*}[!h]
\vspace{-10pt}
\centering
\small
 \setlength{\tabcolsep}{4pt}{
\begin{tabular}{l | lllll | lllll}
  \bottomrule
& \multicolumn{5}{c|}{\bf DBLP Networks} & \multicolumn{5}{c}{\bf IMDB Networks}\\
\hline
Models &{\bf LCC}&{\bf TC} &{\bf CPL} &{\bf GINI} & {\bf REDE} &{\bf LCC}&{\bf TC} &{\bf CPL} &{\bf GINI} & {\bf REDE}\\
 \hline
Original & 107.5 & 59.90 & 3.6943 & 0.3248 & 0.9385 & 13.001 & 305.9 & 1.2275 & 0.1222 & 0.9894\\
\hline
GVAE (no DP) & 7.51 & 66.93 &  0.1330 &  0.0213 & 0.0084 & 0.0145 & 25.83 &  0.0121 & 0.0030 &  0.0016 \\
GGAN (no DP) & 7.23 & 56.29 & 0.1293 & 0.0201 & 0.0057 & 0.0040 & 21.71 & 0.0109 & 0.0010 & 0.0012 \\
NetGAN (no DP)& 9.66 & 39.87 & 0.1943 & 0.0105 & 0.0022 & 0.0083 & 27.54 & 0.0192 & 0.0042 & 0.0011 \\
GraphRNN (no DP)& 10.27 & 57.43 & 0.2043 & 0.0415 & 0.0052 & 0.0594 & 27.26 & 0.0214 & 0.0155 & 0.0094 \\
\hline
\back ($\varepsilon$=10) & {\bf 21.96} & 175.29 & {\bf 0.2471} & {\bf 0.0339} & {\bf 0.0153}  & 0.0147 & 43.63 & 0.0367 & 0.0036 & 0.0030 \\
\back ($\varepsilon$=1) & 23.80 & 187.20 & 0.3059 & {\bf 0.0343} & 0.0156 & 0.0253 & 43.73 & 0.0373 & 0.0038 & 0.0031\\
\back ($\varepsilon$=0.1) & 26.07 & 215.13 & 0.3342 & 0.0344 & 0.0158 & 0.0320 & 44.12 & 0.0392 & 0.0042 & 0.0032\\
\hline
\dgg ($\varepsilon$=10) & {\bf 9.24} & {\bf 64.75} & {\bf 0.2035} & {\bf 0.0224} & {\bf 0.0093} & {\bf 0.0040} & {\bf 22.89} & {\bf 0.0164} &  {\bf 0.0010} & {\bf 0.0017}\\
\dgg ($\varepsilon$=1) & {\bf 12.38} & {\bf 70.97} & {\bf 0.2643} & 0.0353 & {\bf 0.0117}& {\bf 0.0053} & {\bf 23.81} & {\bf 0.0168} &  {\bf 0.0029} & {\bf 0.0023}\\
\dgg ($\varepsilon$=0.1) & 24.62 & {\bf 77.41} & 0.2713 &  0.0485 & 0.0191 & {\bf 0.0113} & {\bf 24.91} & {\bf 0.0168} & {\bf 0.0029} & {\bf 0.0025}\\
\toprule
\end{tabular}}
\vspace{-10pt}
\caption{Performance evaluation over compared models regarding a suite of important graph structural statistics. The Original rows include the values of original networks, while the rest rows are the average absolute difference between generated networks by different models and the original networks. Therefore, \textit{smaller values} indicate better capturing of global network structure and thus \textit{better global data utility}. Bold font is used to highlight the top-3 models with edge-DP constraints.}
\vspace{-5pt}
\label{tab:prop}
\end{table*}

\begin{table*}[!h]
\centering
\small
 \setlength{\tabcolsep}{4pt}{
\begin{tabular}{l | lll | lll}
  \bottomrule
& \multicolumn{3}{c|}{\bf DBLP Networks} & \multicolumn{3}{c}{\bf IMDB Networks}\\
\hline
Models &{\bf Degree dist.}&{\bf Motif ct.} &{\bf GIN acc.} &{\bf Degree dist.}&{\bf Motif ct.} &{\bf GIN acc.}\\
\hline
GVAE (no DP)  & {\bf 0.6171}  & 0.4093 & 0.3029 & {\bf 0.5132}       & {\bf 0.4129} & {\bf 0.4698} \\
GGAN (no DP) & {\bf0.6258} & {\bf0.4231} & {\bf0.3374} & {\bf0.5493} & {\bf0.4279} & {\bf0.4800} \\
NetGAN (no DP) & 0.5754  & {\bf 0.4109} & {\bf 0.3471}  & 0.4921       & 0.3891 & 0.4350 \\
GraphRNN (no DP)         & 0.5454  & 0.3672 & 0.3210  & 0.4635       & 0.3721 & 0.3875  \\
\hline 
\back ($\varepsilon$=1)  & 0.5476  & 0.4038 & 0.3043  & 0.5081       & 0.4021 & 0.4625 \\
\dgg ($\varepsilon$=1)   & {\bf 0.6092}  & {\bf 0.4150} & {\bf 0.3261}  & {\bf 0.5486}       & {\bf 0.4150} & {\bf 0.4725} \\
\toprule
\end{tabular}}
\vspace{-10pt}
\caption{Performance evaluation regarding degree distribution, motif counts and GIN accuracy. {\it Larger values} for both cosine similarity and GIN accuracy indicate {\it better graph utility}. Note that the GIN model trained on the original networks achieves 0.3578 and 0.5163 accuracy on the two datasets, which provides the upper bounds for all compared graph generation models. Bold font is used for values ranked top-3.} 
\vspace{-15pt}
\label{tab:cos}
\end{table*}

We conduct two sets of experiments to evaluate the effectiveness of \dgg~in \textit{preserving global network structure} and \textit{protecting individual link privacy}.
All code and data are in the Supplementary Materials accompanying the submission.

\header{Experimental settings.}
To provide a side-to-side comparison between the original networks and generated networks, we use two standard datasets of real-world networks, \ie, DBLP, and IMDB. 
DBLP includes 72 networks of author nodes and co-author links, where the average numbers of nodes and links are 177.2 and 258; IMDB includes 1500 networks of actor/actress nodes and co-star links, with average node and link numbers 13 and 65.9. 
To facilitate a better understanding towards how the graph statistics reflect the global network structure captured by the models, we also provide results of two recent deep network generation methods, \ie, NetGAN \cite{bojchevski2018netgan} and GraphRNN \cite{you2018graphrnn}, with default parameter settings and no DP constraints at all. 

Due to space limitation, detailed settings of the neural architectures and hyper-parameters of our models as well as runtime comparison among algorithms are put into Appendix C. In Appendix E, we also provide graph visualizations for qualitative visual inspections.

\vspace{-1mm}
\header{Preserving global structures.}
To show that \dgg~effectively captures global network structures, we compare it and \back~under different privacy budgets (controlled by $\varepsilon$ in Eq.~\eqref{eq:noise}), regarding a suite of graph statistics commonly used to evaluate the performance of graph generation models, especially from a global perspective \cite{bojchevski2018netgan,you2018graphrnn,yang2019conditional}. 
\footnote{Statistics we use including LCC (size of the largest connected component), TC (triangle count), CPL (characteristic path length), GINI (gini index) and REDE (relative edge distribution entropy).}
In particular, we train all models from scratch to convergence on each graph in the datasets. 
Each time, the trained model is used to generate one graph, which is compared with the original one regarding the suite of graph statistics. Then we average the absolute differences between generated and original graphs, ensuring that the positive and negative differences do not cancel out. 

Beyond the single value statistics, we also compare the generated graph regarding degree distribution and motif counts. For degree distribution, we convert each graph into a 50-dim vector (all nodes with degree larger than 50 are binned together); for motif counts, we enumerate all 29 undirected motifs with 3-5 nodes and convert each graph into a 29-dim vector by motif matching. We compute the average cosine similarity between pairs of original graphs and generated graphs. 
Furthermore, we use graph classification, the most widely studied graph-level downstream task, to evaluate the global utilities of generated graphs. Particularly, we evaluate the accuracy of GIN \cite{xu2018powerful}, the state-of-the-art graph classification model, with the default parameter setting.


In Table \ref{tab:prop}, our strictly DP-constrained models constantly yield highly competitive and even better results compared with the non-private baselines of NetGAN and GraphRNN regarding global graph structural similarity between generated and original networks on both datasets.
As we gradually increase the privacy budget $\varepsilon$, our both models apparently perform better, showing the effectiveness of our privacy constraints and a clear trade-off between privacy and utility. 
\dgg~consistently outperforms \back~ under the same privacy budgets, supporting our novel design of the GAN framework.
Moreover, as shown in Table \ref{tab:cos}, the graphs generated by \dgg~are competitively similar to the original graphs regarding both degree distributions and motif counts, while achieving satisfactory graph classification accuracy.
All these results consistently demonstrate the global structure preservation ability of \dgg.

\begin{figure}[h]
\centering
\vspace{-3mm}
\includegraphics[width=0.45\textwidth]{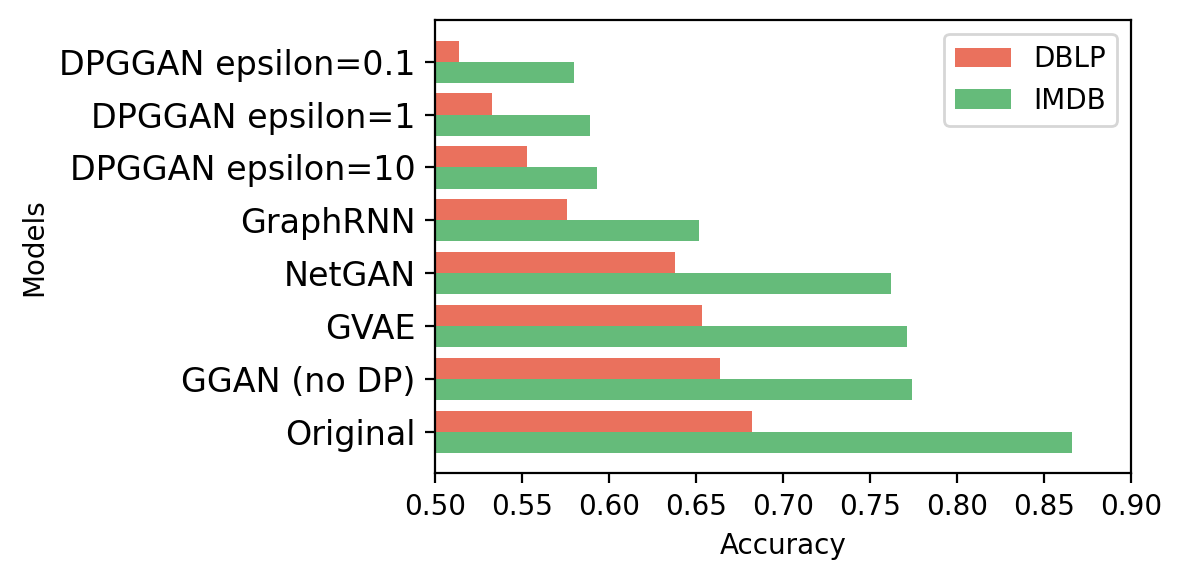}
\vspace{-3mm}
\caption{Accuracy of links predicted on networks generated by \dgg~with varying $\varepsilon$ values compared with baselines. \textit{Lower accuracy} means \textit{better individual link privacy}.}
\label{fig:lp}
\vspace{-2.5mm}
 \end{figure}

\vspace{-2mm}
\header{Protecting individual links.}
To show that \dgg~effectively guarantees individual link privacy, we train models shown in Figure~\ref{fig:lp} for another $K$ times on each dataset. Instead of complete networks, we randomly sample 80\% of the original networks' links to train the models. After training and generation, we use degree distribution to align the nodes in the generated networks with those in the original networks. Then we evaluate the individual link prediction accuracy 
by comparing links predicted in the generated networks and links hidden during training in the original networks.


As shown in Figure~\ref{fig:lp}, for both datasets, links predicted on the networks generated by \dgg~models are much \textit{less accurate} than those predicted on the original networks (13.0\%-16.9\% and 27.3\%-28.6\% accuracy drops on DBLP and IMDB, respectively) as well as the networks generated by most non-private models (11.1\%-15\% and 18.1\%-19.5\% accuracy drops compared with GGAN on DBLP and IMDB, respectively). This means even if the attackers identify nodes in the generated (released) networks of \dgg, they cannot leverage the information there to accurately infer the existence or absence of links between particular pairs of nodes on the original networks. This directly corroborates our claim that \dgg~is effective in protecting individual link privacy. 
Due to space limit, more details and discussions regarding the experimental results are put into Appendix D. 

\vspace{-10pt}
\section{Conclusion}
\vspace{-5pt}
\label{sec:con}
Due to the recent development of deep graph generation models, synthetic networks are generated and released for granted, without the concern about possible privacy leakage over the original networks used for model training. 
In this work, for the first time, we provide a compelling system for secure graph generation through the appropriate integration of deep graph generation models and differential privacy.
Comprehensive experiments show our model to be effective in both preserving global graph structure and protecting individual link privacy.

\small

\bibliographystyle{named}
\bibliography{carlyang} 

\begin{thebibliography}{}

\bibitem[\protect\citeauthoryear{Abadi \bgroup \em et al.\egroup
  }{2016}]{Abadi:2016:DLD:2976749.2978318}
Martin Abadi, Andy Chu, Ian Goodfellow, H.~Brendan McMahan, Ilya Mironov, Kunal
  Talwar, and Li~Zhang.
\newblock Deep learning with differential privacy.
\newblock In {\em SIGSAC}, 2016.

\bibitem[\protect\citeauthoryear{Barab{\'a}si and
  Albert}{1999}]{barabasi1999emergence}
Albert-L{\'a}szl{\'o} Barab{\'a}si and R{\'e}ka Albert.
\newblock Emergence of scaling in random networks.
\newblock {\em science}, 1999.

\bibitem[\protect\citeauthoryear{Blocki \bgroup \em et al.\egroup
  }{2012}]{Blocki12EdgeDP}
Jeremiah Blocki, Avrim Blum, Anupam Datta, and Or~Sheffet.
\newblock The johnson-lindenstrauss transform itself preserves differential
  privacy.
\newblock {\em FOCS}, 2012.

\bibitem[\protect\citeauthoryear{Bojchevski \bgroup \em et al.\egroup
  }{2018}]{bojchevski2018netgan}
Aleksandar Bojchevski, Oleksandr Shchur, Daniel Z{\"u}gner, and Stephan
  G{\"u}nnemann.
\newblock Netgan: Generating graphs via random walks.
\newblock In {\em ICML}, 2018.

\bibitem[\protect\citeauthoryear{{Cai} \bgroup \em et al.\egroup
  }{2018}]{cai2018collective}
Z.~{Cai}, Z.~{He}, X.~{Guan}, and Y.~{Li}.
\newblock Collective data-sanitization for preventing sensitive information
  inference attacks in social networks.
\newblock {\em TDSC}, 2018.

\bibitem[\protect\citeauthoryear{Dwork \bgroup \em et al.\egroup
  }{2014}]{dwork2014algorithmic}
Cynthia Dwork, Aaron Roth, et~al.
\newblock The algorithmic foundations of differential privacy.
\newblock {\em Theoretical Computer Science}, 9(3--4):211--407, 2014.

\bibitem[\protect\citeauthoryear{Fredrikson \bgroup \em et al.\egroup
  }{2015}]{fredrikson2015model}
Matt Fredrikson, Somesh Jha, and Thomas Ristenpart.
\newblock Model inversion attacks that exploit confidence information and basic
  countermeasures.
\newblock In {\em SIGSAC}, 2015.

\bibitem[\protect\citeauthoryear{Frigerio \bgroup \em et al.\egroup
  }{2019}]{frigerio2019differentially}
Lorenzo Frigerio, Anderson~Santana de~Oliveira, Laurent Gomez, and Patrick
  Duverger.
\newblock Differentially private generative adversarial networks for time
  series, continuous, and discrete open data.
\newblock In {\em IFIP SEC}, 2019.

\bibitem[\protect\citeauthoryear{Gu \bgroup \em et al.\egroup
  }{2019}]{gu2018dialogwae}
Xiaodong Gu, Kyunghyun Cho, Jungwoo Ha, and Sunghun Kim.
\newblock Dialogwae: Multimodal response generation with conditional
  wasserstein auto-encoder.
\newblock In {\em ICLR}, 2019.

\bibitem[\protect\citeauthoryear{Hamilton \bgroup \em et al.\egroup
  }{2017}]{hamilton2017inductive}
Will Hamilton, Zhitao Ying, and Jure Leskovec.
\newblock Inductive representation learning on large graphs.
\newblock In {\em NIPS}, 2017.

\bibitem[\protect\citeauthoryear{Hu \bgroup \em et al.\egroup
  }{2020}]{hu2020open}
Weihua Hu, Matthias Fey, Marinka Zitnik, Yuxiao Dong, Hongyu Ren, Bowen Liu,
  Michele Catasta, and Jure Leskovec.
\newblock Open graph benchmark: Datasets for machine learning on graphs.
\newblock {\em arXiv preprint arXiv:2005.00687}, 2020.

\bibitem[\protect\citeauthoryear{Kasiviswanathan \bgroup \em et al.\egroup
  }{2013}]{Kasiviswanathan13NodeDP}
Shiva~Prasad Kasiviswanathan, Kobbi Nissim, Sofya Raskhodnikova, and Adam~D.
  Smith.
\newblock Analyzing graphs with node differential privacy.
\newblock In {\em TCC}, 2013.

\bibitem[\protect\citeauthoryear{Kipf and Welling}{2016}]{kipf2016variational}
Thomas~N Kipf and Max Welling.
\newblock Variational graph auto-encoders.
\newblock In {\em NIPS Workshop on Bayesian Deep Learning}, 2016.

\bibitem[\protect\citeauthoryear{Kipf and Welling}{2017}]{kipf2016semi}
Thomas~N Kipf and Max Welling.
\newblock Semi-supervised classification with graph convolutional networks.
\newblock In {\em ICLR}, 2017.

\bibitem[\protect\citeauthoryear{Kuhn}{2009}]{kuhn2009compensation}
Kristine~M Kuhn.
\newblock Compensation as a signal of organizational culture: the effects of
  advertising individual or collective incentives.
\newblock {\em IJHRM}, 20(7):1634--1648, 2009.

\bibitem[\protect\citeauthoryear{Kurakin \bgroup \em et al.\egroup
  }{2017}]{kurakin2016adversarial}
Alexey Kurakin, Ian Goodfellow, and Samy Bengio.
\newblock Adversarial machine learning at scale.
\newblock {\em ICLR}, 2017.

\bibitem[\protect\citeauthoryear{Maron \bgroup \em et al.\egroup
  }{2019}]{maron2019invariant}
Haggai Maron, Heli Ben-Hamu, Nadav Sharmir, and Lipman Yaron.
\newblock Invariant and equivariant graph networks.
\newblock In {\em ICLR}, 2019.

\bibitem[\protect\citeauthoryear{Narayanan and
  Shmatikov}{2009}]{narayanan2009anonymizing}
Arvind Narayanan and Vitaly Shmatikov.
\newblock De-anonymizing social networks.
\newblock {\em SP}, 2009.

\bibitem[\protect\citeauthoryear{Nobari \bgroup \em et al.\egroup
  }{2014}]{nobari2014opacity}
Sadegh Nobari, Panagiotis Karras, Hwee~Hwa PANG, and St{\'e}phane Bressan.
\newblock L-opacity: linkage-aware graph anonymization.
\newblock 2014.

\bibitem[\protect\citeauthoryear{Papernot \bgroup \em et al.\egroup
  }{2018}]{papernot2018scalable}
Nicolas Papernot, Shuang Song, Ilya Mironov, Ananth Raghunathan, Kunal Talwar,
  and {\'U}lfar Erlingsson.
\newblock Scalable private learning with pate.
\newblock In {\em ICLR}, 2018.

\bibitem[\protect\citeauthoryear{Sala \bgroup \em et al.\egroup
  }{2011}]{Sala11ShareGraphDP}
Alessandra Sala, Xiaohan Zhao, Christo Wilson, Haitao Zheng, and Ben~Y. Zhao.
\newblock Sharing graphs using differentially private graph models.
\newblock In {\em SIGCOMM}, 2011.

\bibitem[\protect\citeauthoryear{Shokri \bgroup \em et al.\egroup
  }{2017}]{shokri2017membership}
Reza Shokri, Marco Stronati, Congzheng Song, and Vitaly Shmatikov.
\newblock Membership inference attacks against machine learning models.
\newblock In {\em ISP}, 2017.

\bibitem[\protect\citeauthoryear{Sigurbj{\"o}rnsson and
  Van~Zwol}{2008}]{sigurbjornsson2008flickr}
B{\"o}rkur Sigurbj{\"o}rnsson and Roelof Van~Zwol.
\newblock Flickr tag recommendation based on collective knowledge.
\newblock In {\em WWW}, 2008.

\bibitem[\protect\citeauthoryear{Simonovsky and
  Komodakis}{2018}]{simonovsky2018graphvae}
Martin Simonovsky and Nikos Komodakis.
\newblock Graphvae: Towards generation of small graphs using variational
  autoencoders.
\newblock In {\em ICANN}, 2018.

\bibitem[\protect\citeauthoryear{Sun and Lyu}{2020}]{sun2020federated}
Lichao Sun and Lingjuan Lyu.
\newblock Federated model distillation with noise-free differential privacy.
\newblock {\em arXiv preprint arXiv:2009.05537}, 2020.

\bibitem[\protect\citeauthoryear{Sun \bgroup \em et al.\egroup
  }{2018}]{sun2018adversarial}
Lichao Sun, Yingtong Dou, Carl Yang, Ji~Wang, Philip~S Yu, Lifang He, and
  Bo~Li.
\newblock Adversarial attack and defense on graph data: A survey.
\newblock {\em arXiv preprint arXiv:1812.10528}, 2018.

\bibitem[\protect\citeauthoryear{Sun \bgroup \em et al.\egroup
  }{2020a}]{sun2020ldp}
Lichao Sun, Jianwei Qian, Xun Chen, and Philip~S Yu.
\newblock Ldp-fl: Practical private aggregation in federated learning with
  local differential privacy.
\newblock {\em arXiv preprint arXiv:2007.15789}, 2020.

\bibitem[\protect\citeauthoryear{Sun \bgroup \em et al.\egroup
  }{2020b}]{sun2020differentially}
Lichao Sun, Yingbo Zhou, Philip~S Yu, and Caiming Xiong.
\newblock Differentially private deep learning with smooth sensitivity.
\newblock {\em arXiv preprint arXiv:2003.00505}, 2020.

\bibitem[\protect\citeauthoryear{Wang and
  Wu}{2013}]{Wang13DPDegreeGraphGeneration}
Yue Wang and Xintao Wu.
\newblock Preserving differential privacy in degree-correlation based graph
  generation.
\newblock {\em TDP}, 2013.

\bibitem[\protect\citeauthoryear{Watts and
  Strogatz}{1998}]{watts1998collective}
Duncan~J Watts and Steven~H Strogatz.
\newblock Collective dynamics of `small-world' networks.
\newblock {\em nature}, 393(6684):440, 1998.

\bibitem[\protect\citeauthoryear{Xie \bgroup \em et al.\egroup
  }{2020}]{xie2020gnns}
Yiqing Xie, Sha Li, Carl Yang, Raymond Chi-Wing Wong, and Jiawei Han.
\newblock When do gnns work: Understanding and improving neighborhood
  aggregation.
\newblock In {\em IJCAI}, 2020.

\bibitem[\protect\citeauthoryear{Xu \bgroup \em et al.\egroup
  }{2019}]{xu2018powerful}
Keyulu Xu, Weihua Hu, Jure Leskovec, and Stefanie Jegelka.
\newblock How powerful are graph neural networks?
\newblock In {\em ICLR}, 2019.

\bibitem[\protect\citeauthoryear{Xue \bgroup \em et al.\egroup
  }{2012}]{xue2012delineating}
Mingqiang Xue, Panagiotis Karras, Raissi Chedy, Panos Kalnis, and Hung~Keng
  Pung.
\newblock Delineating social network data anonymization via random edge
  perturbation.
\newblock In {\em CIKM}, 2012.

\bibitem[\protect\citeauthoryear{Yang \bgroup \em et al.\egroup
  }{2019}]{yang2019conditional}
Carl Yang, Peiye Zhuang, Wenhan Shi, Alan Luu, and Pan Li.
\newblock Conditional structure generation through graph variational generative
  adversarial nets.
\newblock In {\em NIPS}, 2019.

\bibitem[\protect\citeauthoryear{You \bgroup \em et al.\egroup
  }{2018}]{you2018graphrnn}
Jiaxuan You, Rex Ying, Xiang Ren, William~L Hamilton, and Jure Leskovec.
\newblock Graphrnn: Generating realistic graphs with deep auto-regressive
  models.
\newblock In {\em ICML}, 2018.

\bibitem[\protect\citeauthoryear{Zhang \bgroup \em et al.\egroup
  }{2014}]{zhang2014privacy}
Aston Zhang, Xing Xie, Kevin Chen-Chuan Chang, Carl~A Gunter, Jiawei Han, and
  XiaoFeng Wang.
\newblock Privacy risk in anonymized heterogeneous information networks.
\newblock In {\em EDBT}, 2014.

\bibitem[\protect\citeauthoryear{Zhang \bgroup \em et al.\egroup
  }{2019}]{zhang2019enabling}
Yanjun Zhang, Xin Zhao, Xue Li, Mingyang Zhong, Caitlin Curtis, and Chen Chen.
\newblock Enabling privacy-preserving sharing of genomic data for gwass in
  decentralized networks.
\newblock In {\em WSDM}, 2019.

\end{thebibliography}

\end{document}